\documentclass[12pt,a4paper]{article}
\usepackage{indentfirst}
\usepackage{times}
\usepackage[font=small,labelfont=bf]{caption} 
\usepackage{amsfonts, amsmath, amsthm, amssymb} 
\usepackage{wrapfig} 
\usepackage{graphicx} 
\usepackage[utf8]{inputenc}
\usepackage[english]{babel}
\usepackage{setspace}
\usepackage{authblk}
\usepackage{color}
\usepackage[symbol*]{footmisc}
\usepackage{url} 
\usepackage{wrapfig}
\usepackage[noadjust]{cite}
\usepackage{mdframed}

\graphicspath{{}}
\begin{document}
\sloppy
\noindent\textbf{\Large Ion-irradiated YBa$_2$Cu$_3$O$_7$ Josephson arrays}\par
\vspace{10pt}
\noindent A.~Sharafiev$^{1}$\footnote[1]{sharafiev@physics.msu.ru}, M.~Malnou$^1$, C.~Feuillet-Palma$^1$, C.~Ulysse$^2$, P.~Febvre$^3$, J.~Lesueur$^1$, N.~Bergeal$^1$\par
\vspace{7pt}
\noindent\scriptsize$^1$Laboratoire de Physique et d'Etude des Mat\'eriaux - UMR8213-CNRS-ESPCI Paris-UPMC, PSL Research University,10 Rue Vauquelin - 75005 Paris, France.\\$^2$Laboratoire de Photonique et de Nanostructures LPN-CNRS, Route de Nozay, 91460 Marcoussis, France.
\\$^3$IMEP-LAHC - UMR 5130 CNRS, Universit\'e Savoie Mont Blanc, 73376 Le Bourget du Lac cedex, France.\par

\renewcommand{\thefootnote}{\arabic{footnote}}

\vspace{10pt}
\noindent\small\textbf{Abstract.}
We designed, fabricated and tested short one dimensional arrays of masked ion-irradiated YBa$_2$Cu$_3$O$_7$ Josephson junctions (JJ) embedded into log-periodic spiral antennas. Our arrays consist of 4 or 8 junctions separated either by 960~nm or 80~nm long areas of undamaged YBCO. Samples with distanced junctions and with closely spaced junctions showed qualitatively different behaviors. Well separated arrays demonstrated giant Shapiro steps in the hundreds-GHz band at 66K and were tested as Josephson mixers with improved impedance matching. All closely spaced arrays behaved as one junction with a lower superconducting transition temperature, hence forming a single weak link on distances up to 880~nm. Such design opens a new way to increase the I$_{c}$R$_{N}$ product of ion-irradiated junctions and we speculate that the phenomena and physics behind it might be similar to the so-called "giant" Josephson coupling observed in cuprates.    
\normalsize
\section{Introduction}
Ion-irradiation technique is a promising approach to manufacturing high temperature superconducting integrated circuits \cite{katz1998,bergeal2005}. It allows designing a large number of arbitrary located Josephson junctions (JJ) on a single superconducting film and therefore offers a natural scalability compared to other types of high-T$_c$ junctions. The technique was first demonstrated by Tinchev \cite{tinchev1990} for rf SQUID experiments and since then many other applications were suggested: dc SQUIDs \cite{bergeal2006}, SQIFs \cite{ouanani2016}, Josephson mixers \cite{malnou2012,malnou2014} and even digital circuits \cite{wolf2013}. Study of properties of ion-irradiated junctions and further development of the technology attracted considerable attention during the last years. Recent researches  in the area focused on ion-irradiated MgB$_2$ junctions \cite{cybart2006}, direct junction patterning with Focused Ion Beam Irradiation (FIBI) technology \cite{cybart2015,cho2015} and study of temperature dependence of Josephson generation linewidth \cite{sharafiev2016}.\par 
At the same time one of the main drawbacks of the technology restricting its applicability in real systems is the difficulty to produce high-impedance junctions. Normal resistance of single ion-irradiated junctions is typically of order of a few ohms which makes it difficult to couple it with standard 50-ohm microwave environment. In particular, it has been identified as one of the main issues for mixing applications where only moderate conversion efficiency has been demonstrated \cite{malnou2012,malnou2014}. One of possible ways to overcome the problem consists in replacing single JJ with an array of coherently operating junctions  in combination with planar coupling structures. The interest in Josephson arrays originates from theoretical considerations (\cite{jain1984} and associated references) which have shown potential superiority of arrays based devices over single-junction ones. Operating prototypes of array-based devices for different applications have been developed: voltage standards \cite{benz1996,klushin1996}, wave form generators \cite{benz1998}, detectors and mixers \cite{konopka1995,tsuru1995}, etc. \par
Ion irradiation technology is capable of delivering reliable arrays of large scale \cite{chen2004,cybart2008,cybart2009,ouanani2016}. For example in the recent reference {\cite{cybart2014} serial SQIF-sensor consisting of 36000 junctions has been reported. Junctions uniformity is sufficient to obtain giant Shapiro steps \cite{chen2004,cybart2005}.\par
One unique property of the irradiation technology lies into the possibility to fabricate very densely packed arrays keeping it lumped while benefiting at the same time from planar geometry (unlike e.g. stacked or intrinsic junctions, see review in \cite{seidel_handbook}). Moreover, since many applications require a coherent operation of all junctions in the array, external resonators are usually employed to provide long-range synchronization mechanism (theoretical review \cite{hansen1984}, experimental results \cite{barbara1999,han1994,constantinian2000,klushin1996,klushin2008}). In turn, short-range interaction is almost never discussed in the context of planar structures because of technical difficulty of dense packing (except for low-T$_c$ microbridges, see \cite{hansen1984} and associated references).\par        
In this article, we report experimental results on short ion-irradiated Josephson arrays, compare and study their behavior depending on temperature and spacing between junctions. In particular, we address the question of mixing on giant Shapiro steps with an array of well separated junctions and also discuss the physics of closely spaced arrays.
\begin{figure}[!t]
\includegraphics[width=0.45\linewidth]{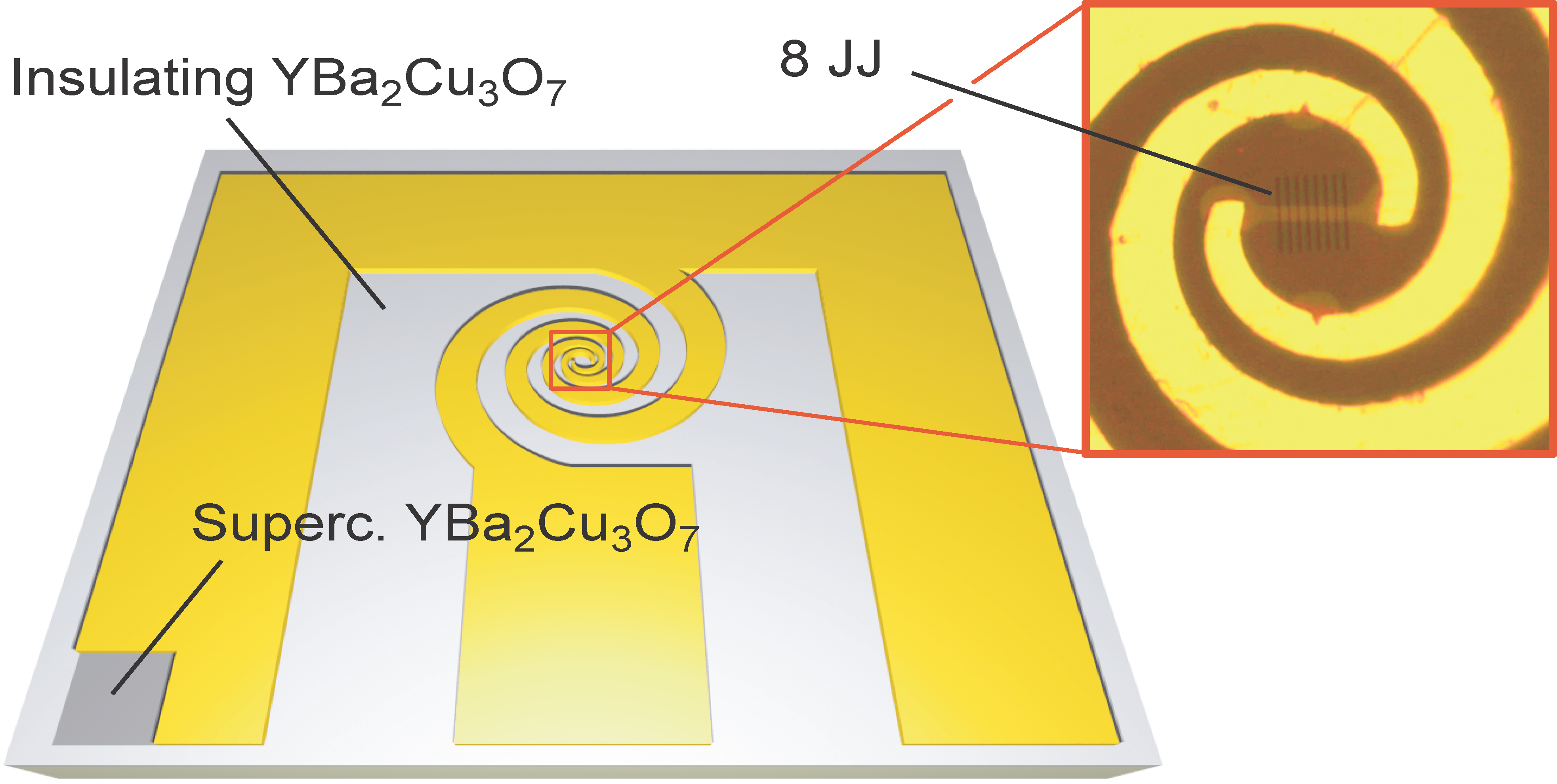}\hspace{1cm}
\includegraphics[width=0.45\linewidth]{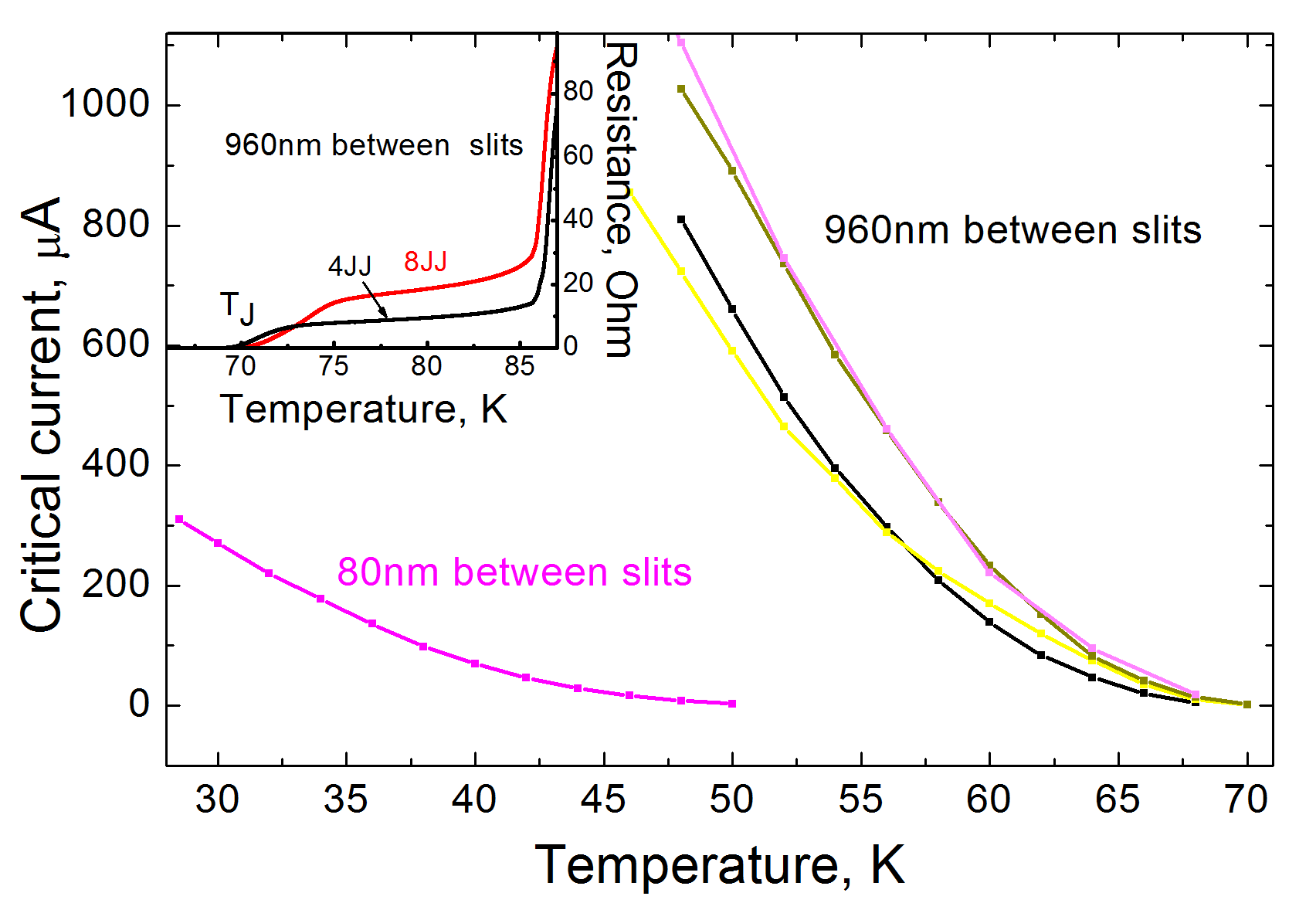}\par
\hspace{4.3cm}\textit{(a)}\hspace{8.2cm}\textit{(b)}
\captionof{figure}{\textit{(a)} Schematic image of the sample with photo of its central part including 2~$\mu$m-bridge with 8 serial Josephson junctions. The external turn diameter of the antenna is $\sim$1mm; \textit{(b)} Typical temperature dependence of the critical current: four samples with 960~nm-spaced junctions (on the right side of the plot) and one with 80~nm spaced (on the left side). Inset: typical temperature dependence of the resistance of 960~nm spaced arrays consisting of 4 and 8 junctions.}  
\label{fig1}
\vspace{-0.5cm}
\end{figure}   
\section{Experimental}
For optimal coupling with external signals and possibility to read rf signal from the sample all our arrays were embedded into spiral log-periodic antennas and a 50~$\Omega$ coplanar waveguide (CPW) transmission lines (figure~1(a)). Details of our standard fabrication technique can be found in \cite{bergeal2007,malnou2014}. In short, we start from a commercial 70~nm thick YBa$_2$Cu$_3$O$_7$ films grown on a Al$_2$O$_3$ substrate\footnote[2]{Ceraco gmbh.}, and covered with a 200~nm gold layer. The spiral antenna and the CPW transmission line are first defined in the gold layer through a MAN e-beam patterned resist, followed by a 500-eV Ar$^+$ Ion Beam Etching. Then a 2~$\mu$m wide channel located at the center of the antenna is patterned in a MAN e-beam resist, followed by a 70~keV oxygen ion irradiation at a dose of 2$\cdot$10$^{15}$~at/cm$^2$. This process ensures that the regions of the film which are not protected either by the resist or by the gold layer become deeply insulating. Finally, the junction array is defined at the center of the superconducting channel by irradiating through 40-nm wide slits patterned in a PMMA resist, with 110~keV oxygen ions at a dose of 3$\cdot$10$^{13}$~at/cm$^2$.\par
The samples were placed inside a closed cycle refrigerator, at the focal point of a Winston cone, and then exposed to the external signals through an optical window. CPW lines of the samples were connected to microwave readout lines through PCB. The signals in the 4-8~GHz bandwidth were first amplified at cryogenic temperature by a HEMT low noise amplifier before further amplification at room temperature. An isolator was placed between the sample and the first amplifier to minimize the back action noise. Dc output voltage and ac output power at intermediate frequency P$_{IF}$ were measured as functions of applied bias current.\par 
\begin{figure}[!t]
\includegraphics[width=0.33\linewidth]{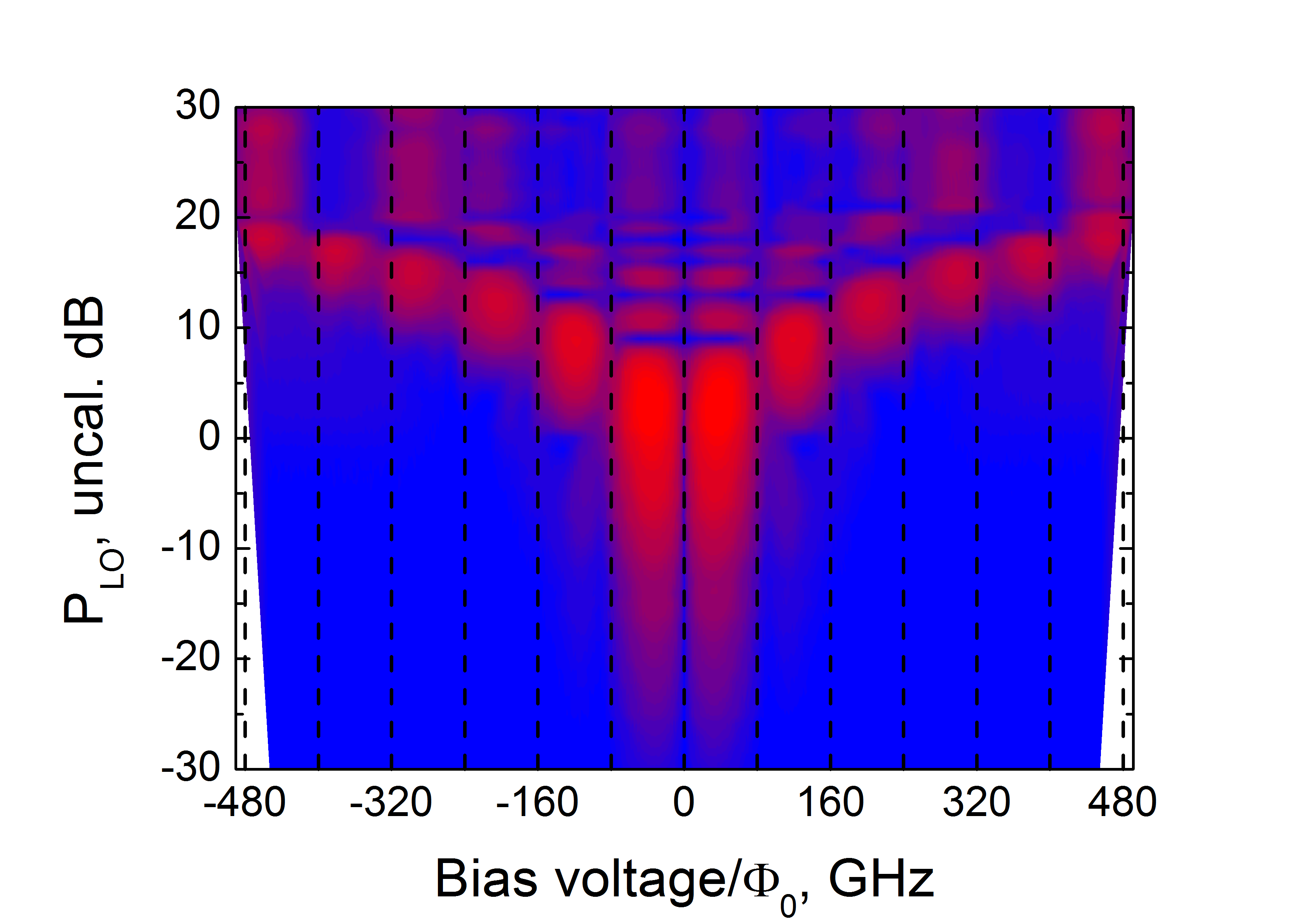}
\includegraphics[width=0.33\linewidth]{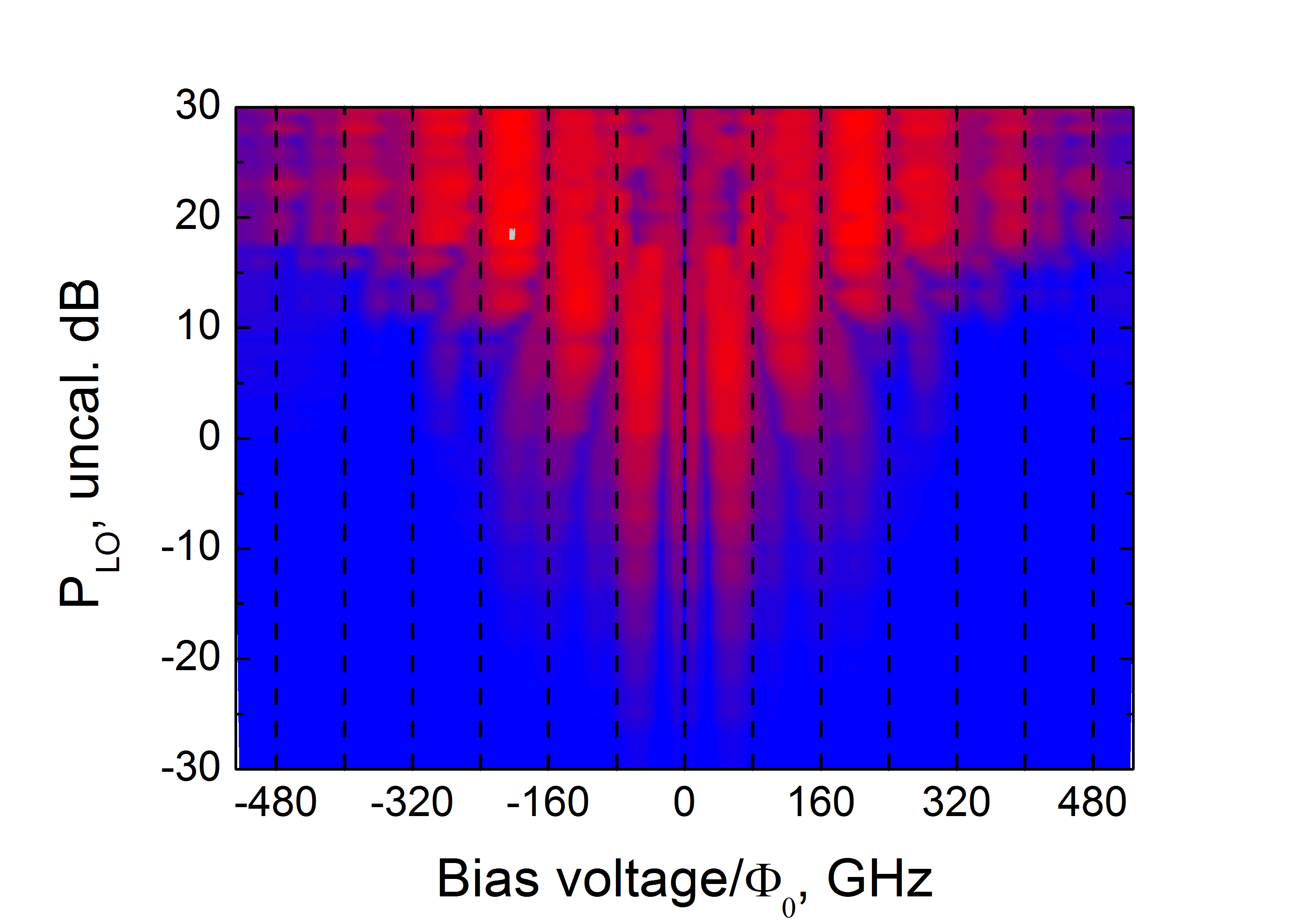}
\includegraphics[width=0.33\linewidth]{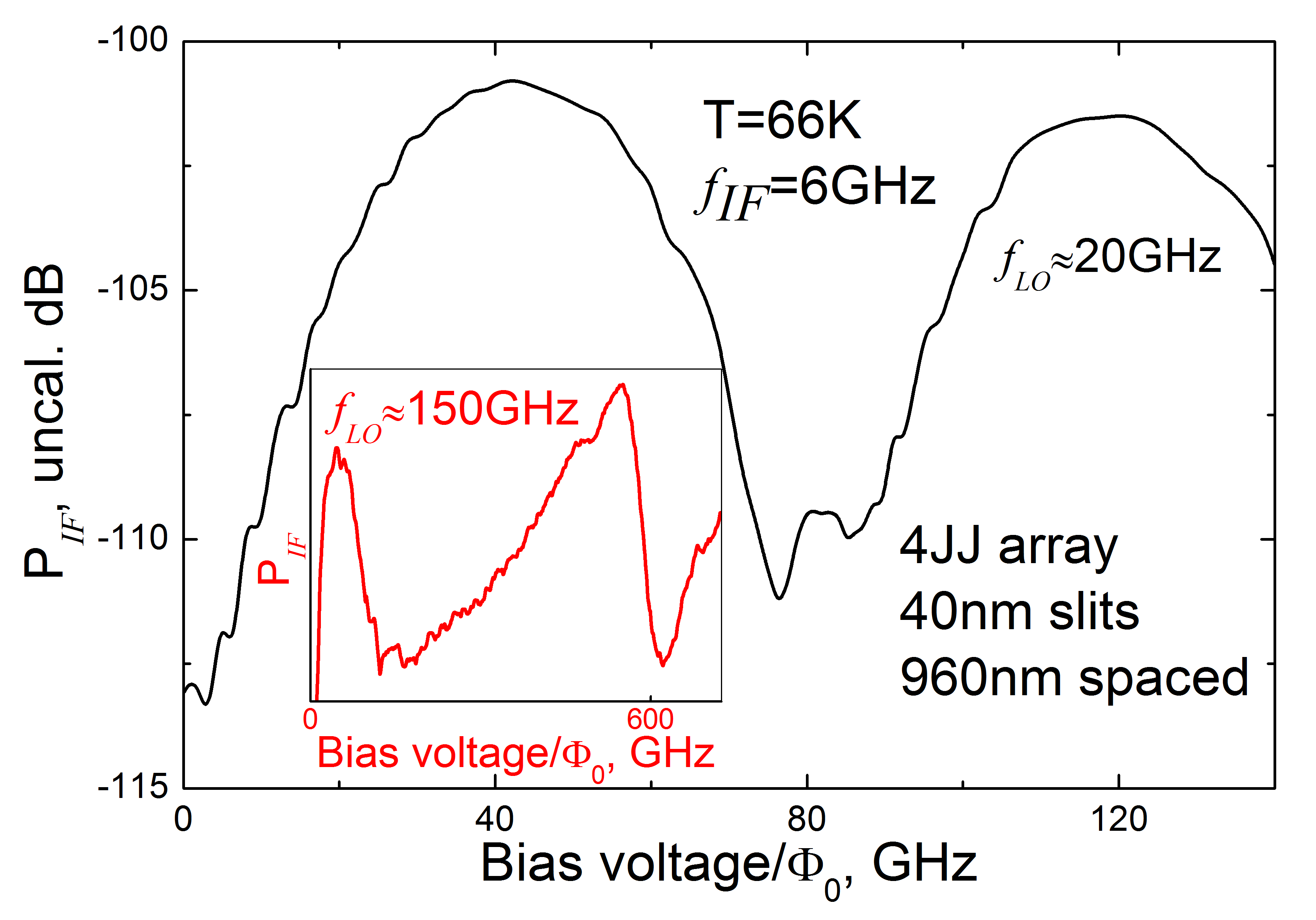}\par
\hspace{3cm}\textit{(a)}\hspace{5cm}\textit{(b)}\hspace{5cm}\textit{(c)}
\captionof{figure}{\textit{(a)} Measured output mixing power P$_{IF}$ at 6GHz (colour, red is for higher P$_{IF}$ and blue is for lower) as a function of local oscillator power P$_{LO}$ and bias voltage for a serial array of 4 ion-irradiated 960~nm spaced junctions at temperature T=66K. Local oscillator frequency $f_{LO}\approx$20GHz, signal frequency $f_{sign}\approx$14GHz. Output power follows differential resistance demonstrating giant Shapiro steps each 4$\cdot$20GHz along bias voltage axis; \textit{(b)} the same for temperature T=60K; \textit{(c)} section of the plot (a) along P$_{LO}$=6dB (black line). Inset: the same for higher frequency mixing ($f_{LO}\approx$150GHz).}  
\label{fig2}
\vspace{-0.5cm}
\end{figure} 
\section{960nm-separated arrays}
This distance between the slits is large enough to ensure that damaged areas do not overlap and that the junctions are independent. All samples demonstrated qualitatively similar behaviors and typical results are shown on figures 1(b), 2 and 3(b). The temperature dependence of the critical currents of the arrays corresponds to the one of a single junction (see curves on the right side of figure 1(b)). For temperatures just below the Josephson coupling  temperature T$_J$ (see the inset in figure 1(b)) the parameters spread among junctions in one array was sufficiently small  to obtain clear giant Shapiro steps (red line on the inset of figure 3(b)). However, further lowering of the temperature increases the parameters spread and instead of giant steps, more chaotic behavior is observed.\par 
This behavior explains the mixing patterns for high (66K) and low (60K) temperatures shown on figure 2 (bias voltage is expressed in frequency units in accordance with Josephson equation $V=f\cdot\Phi_{0}$, $\Phi_{0}$ is flux quantum). In this experiment, a comparatively weak signal with frequency f$_{sign}$=14~GHz was applied to a 4 JJ array along with a local oscillator signal $f_{LO}$=20~GHz whose power was swept. The available power at intermediate frequency P$_{IF}$ was measured with a spectrum analyzer and depicted in colour scale on figures 2(a) and (b). Josephson temperature T$_J$ for this sample was 70~K. At T=66~K the mixing pattern almost coincides with what one would expect from a single junction (figure 2(a) and (c)) but at lower temperatures, the picture becomes more and more fuzzy (figure 2(b) for T=60~K). In accordance with general theory of Josephson mixer output power is proportional to differential resistance (see e.g. \cite{barone_paterno}) and shows either single local maximum between steps when $f_{LO}$ is comparatively low (black line on figure 3(c)) or two maxima and minimum when $f_{LO}$ is higher and steps are well separated (red inset line on the same figure).\par
Josephson mixer properties can be improved by employing JJ array instead of one junction in two ways: better impedance matching and decreasing noise temperature. Sufficient uniformity between junctions (ability to obtain giant Shapiro steps) is necessary to provide equal optimal biasing for all junctions in the array. Providing this condition is satisfied one can fully benefit from better matching and, strictly speaking, mutual phase locking is not a necessity (as it was noticed e.g. in \cite{kuzmin1981}). At the same time, mutual locking would allow to decrease the linewidth of the junctions self-oscillation and therefore decrease the noise temperature which is a substantial problem for Josephson mixers. Indeed, the local oscillator generates additional noise in the output IF band by down conversion of wide Josephson self-oscillations (see e.g. \cite{jain1984}). For the devices directly using Josephson generation to produce output signal (e.g. THz-range generators or devices with internal pumping) mutual phase-locking is a necessity. The fact that short range interaction in some cases can lead to mutual phase locking (\cite{lindelof1977,li1999,kleiner1992,carapella1996}) and that parameters spread in our arrays allows obtaining giant Shapiro steps encouraged us to put slits as close as possible to keep damaged areas still separated but with shorter undamaged areas.            
\begin{figure}[!t]
\includegraphics[width=0.45\linewidth]{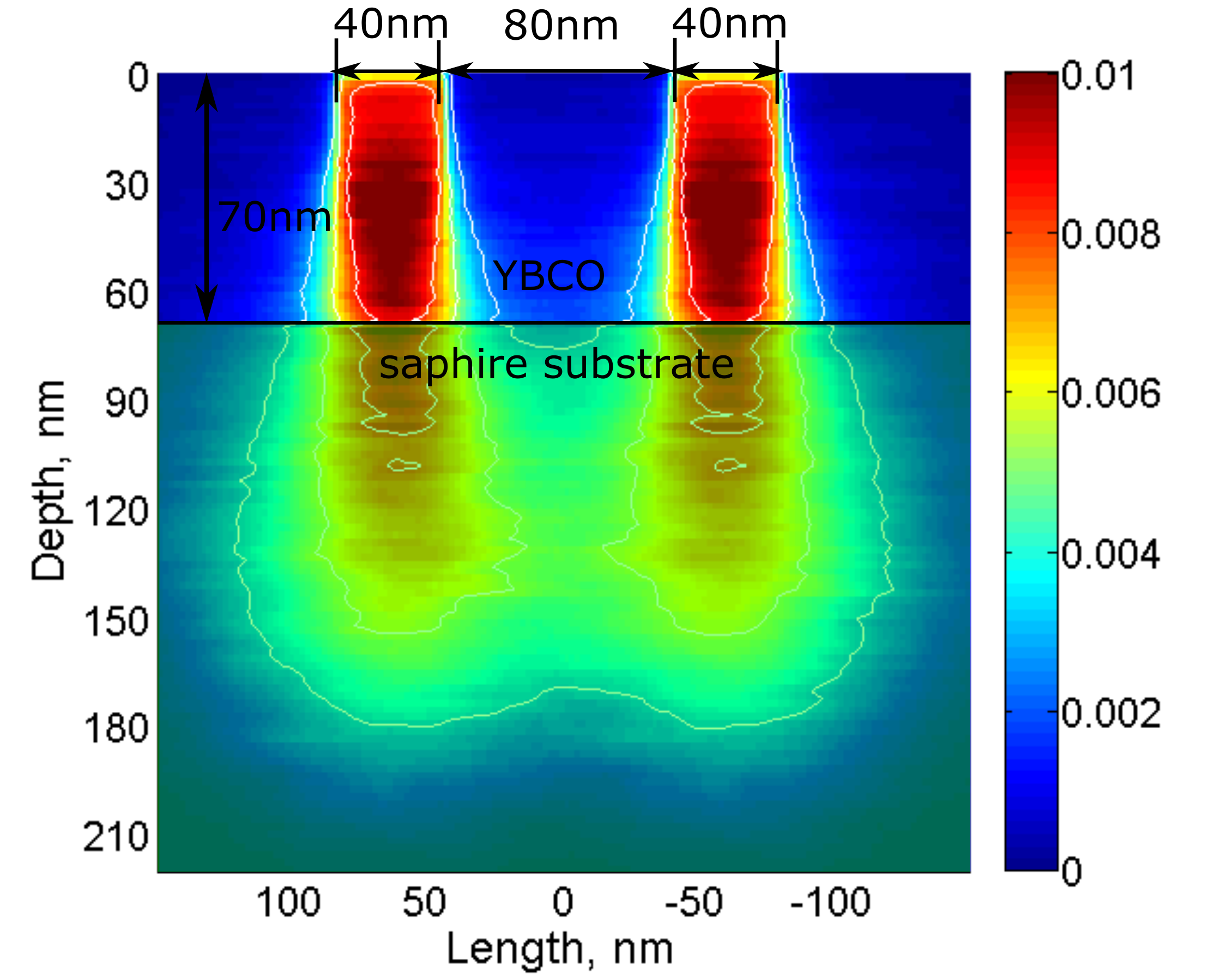}\hspace{1cm}
\includegraphics[width=0.45\linewidth]{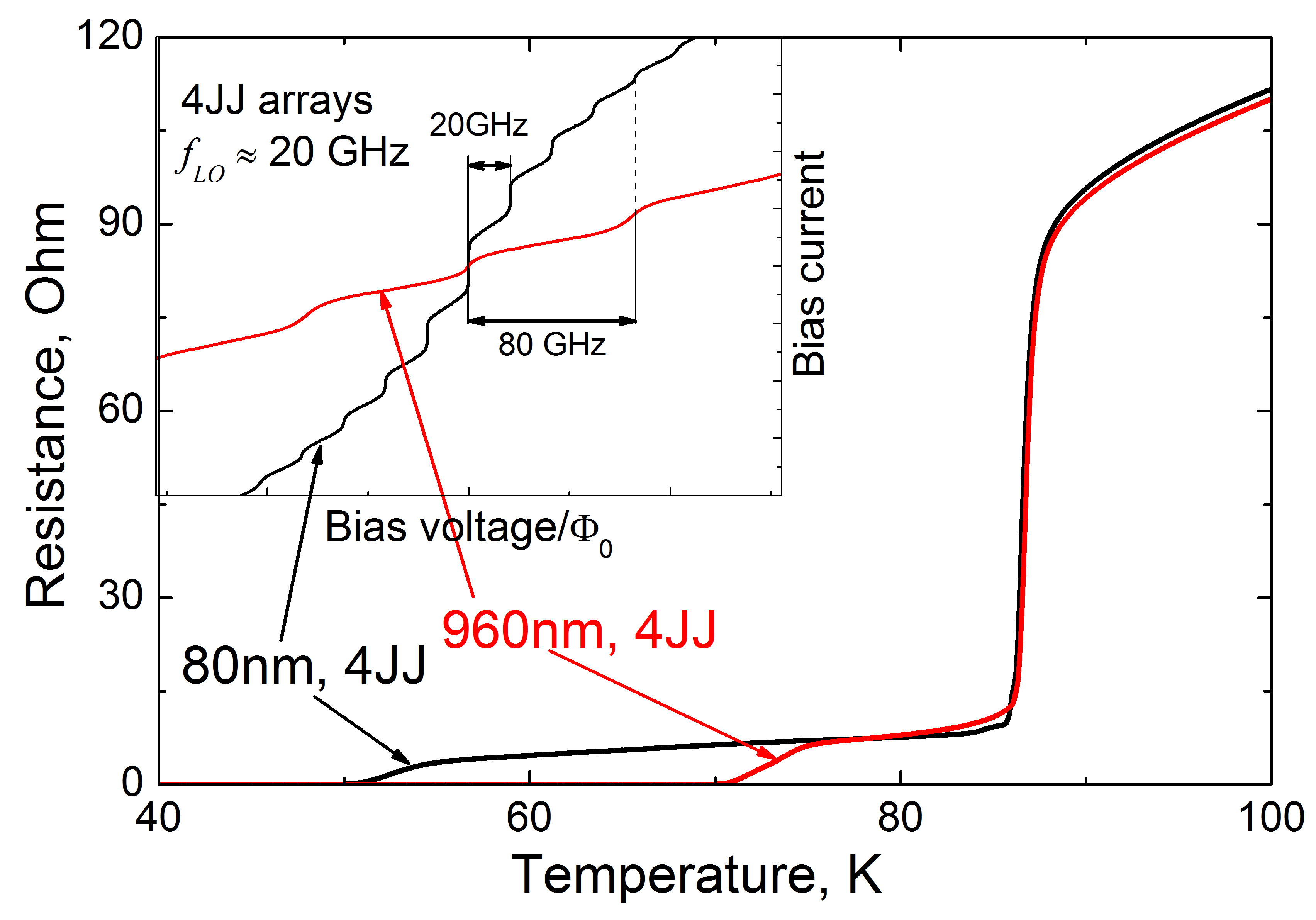}\par
\hspace{4.3cm}\textit{(a)}\hspace{8.2cm}\textit{(b)}
\captionof{figure}{\textit{(a)} Result of Monte-Carlo (SRIM) simulation of O$^+$ irradiation of a 70~nm thick YBa$_2$Cu$_3$O$_7$ microbridge through two 40~nm slits separated by 80~nm of resist-covered area. Colour corresponds to displacements per atom (dpa) level. According to the simulation, two damaged areas almost do not overlap; \textit{(b)} typical temperature dependence of resistivity of 4-slits structures with 960~nm and 80~nm separation. Inset shows Shapiro steps of 80~nm-spaced sample when external signal frequency is 20~GHz and temperature is significantly lower than T$_{J}$. All closely-spaced samples showed single-junction behavior demonstrating Josephson coupling on distances as long as 860~nm.}  
\label{fig3}
\vspace{-0.5cm}
\end{figure}
\section{80nm-separated arrays}
In this experiment, the distance between internal edges of neighboring slits in PMMA was 80~nm as shown in figure 3(a). The figure displays the result of Monte-Carlo (SRIM) simulation of lattice damage for two junctions in an array (brighter areas correspond to higher displacement per atom level). Samples with 4 or 8 slits located in the center of micro bridges were used in the study. The rest of geometry as well as all manufacturing parameters were the same as we used for the samples described in the previous section.\par
All sampled demonstrated quite a surprising behavior illustrated on figure 3(b). Normal resistance scaled with number of slits and was essentially the same as for 960~nm-separated arrays. However, Josephson temperature (T$_J$) was significantly lower and the temperature dependence of critical current was less steep (curve on left side of figure 1(b)). Moreover these samples demonstrated sound one-junction behavior as confirmed by Shapiro steps (inset figure 3(b)). The inset shows two IV-curves for distanced and closely spaced arrays of 4 JJ under external 20~GHz-signal. This behavior has been observed for all produced samples over the full temperature range of interest (corresponding to Josephson regime): from T$_J$ ($\sim$50~K) to 28~K. Note that, as the shown IV-curve of closely-spaced array has been measured at low temperature (T=28~K), the Shapiro steps are comparatively sharp (in agreement with our previous results on generation linewidth temperature dependence \cite{sharafiev2016}), and normal resistance R$_N$ is comparatively low.\par
This indicates that Josephson coupling was established in our structures over large distance between two "bulk" undamaged areas (880~nm for 8-junction array). To our knowledge such a distance for Josephson coupling in YBCO has never been reported yet and deserves a separate discussion.\par
\begin{figure}[!t]
\includegraphics[width=0.45\linewidth]{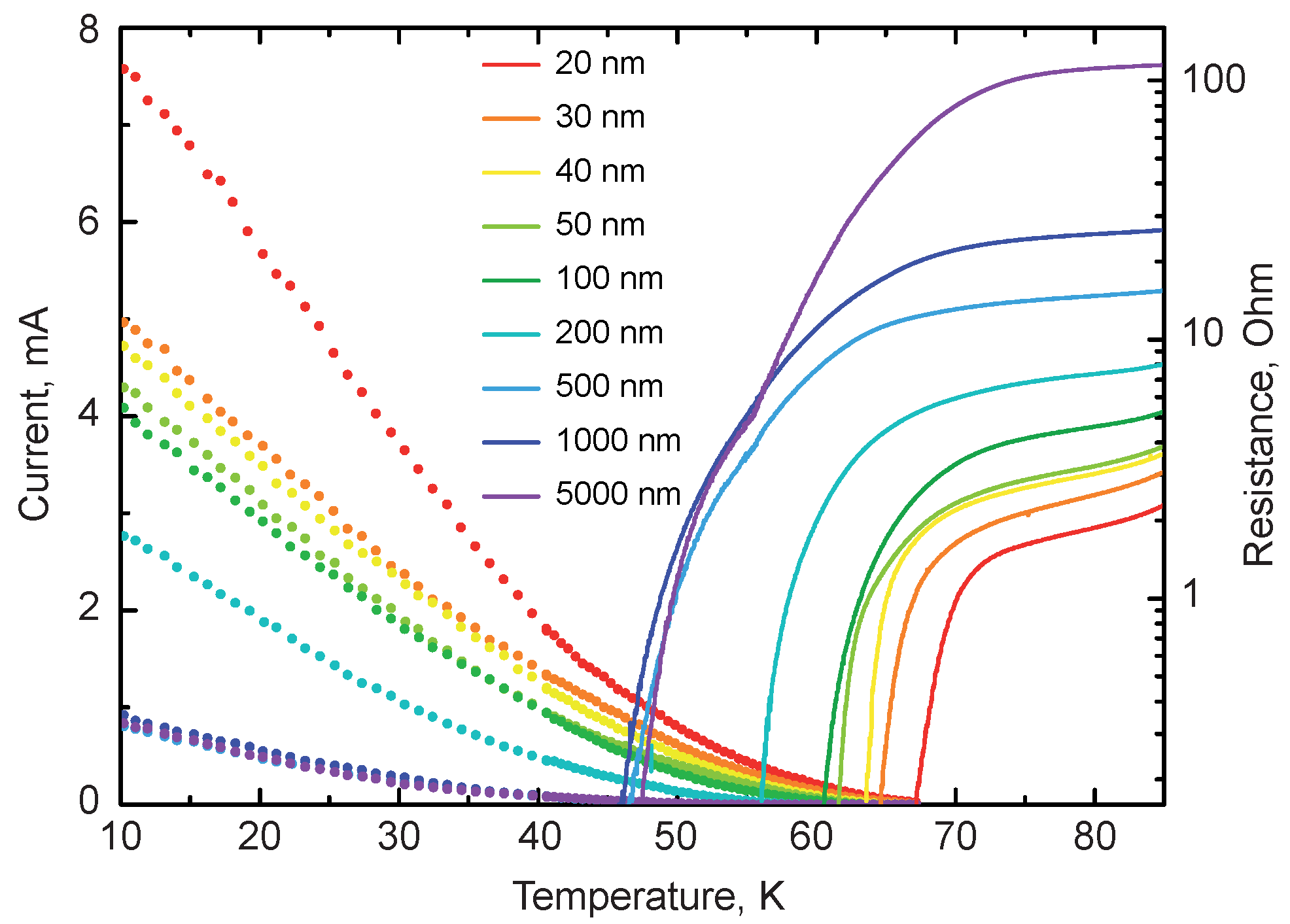}\hspace{1cm}
\includegraphics[width=0.45\linewidth]{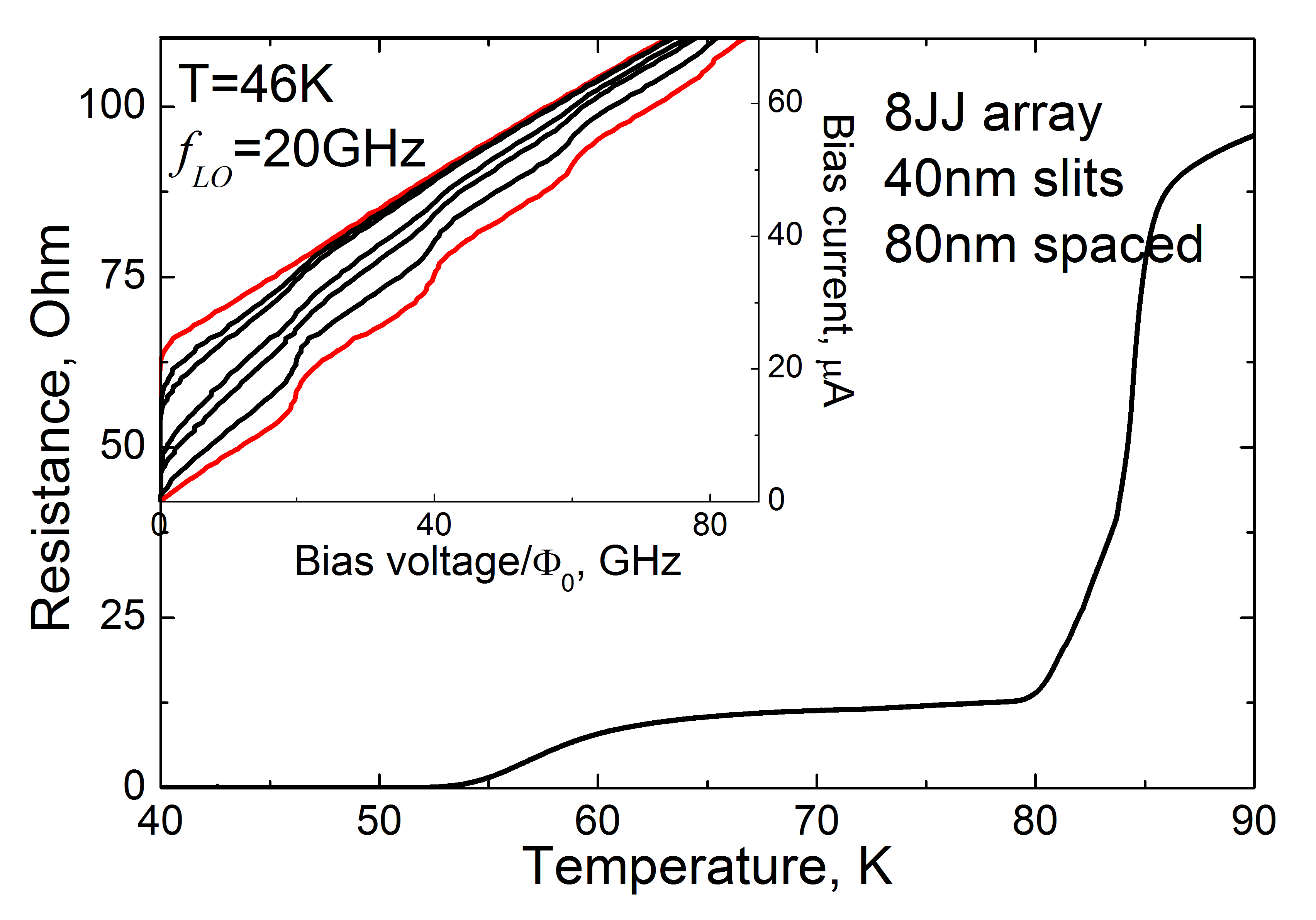}\par
\hspace{4.3cm}\textit{(a)}\hspace{8.2cm}\textit{(b)}
\captionof{figure}{\textit{(a)} R(T) and I$_c$(T) characteristics of YBCO 5~$\mu$m bridges irradiated with dose 3$\cdot$10$^{13}$~ions/cm$^2$ through slits of different width (\cite{wolf_thesis}); \textit{(b)} R(T) and IV-curves (with and without applied 20~GHz external signal) of an array formed by 8 slits spaced with 80~nm of covered areas.}  
\label{fig4}
\vspace{-0.5cm}
\end{figure}   
\section{Discussion}
Ion-irradiated junctions are usually considered as SNS type(\cite{katz1998,katz2000,tinchev1993}) where Josephson coupling is associated with proximity effect, first theoretically described by de Gennes in \cite{de_gennes_1964}. Although this theory was not originally developed for d-wave high-T$_c$ superconductors, it has been exploited many times to describe the different properties of ion-irradiated YBCO junctions (e.g. \cite{katz2000, sharafiev2016}).\par
To our knowledge there have been only two experimental reports on closely spaced locally damaged junctions \cite{booij1997,chen2005} and in both cases, a strong interaction between junctions was observed. Our geometry and manufacturing process is essentially close to the ones described in \cite{chen2005} where authors produced a pair of junctions using masked ion-irradiation technique. By measuring Shapiro steps, they observed the pair operating as a single or two separate junctions depending on the power of applied ac signal.\par
Our previous experiments with 5~$\mu$m wide bridges irradiated through slits of different width with the same dose have shown that the Josephson temperature T$_J$ strongly depends on the width of the slit until it reaches a saturating value (compare lines for 20-200~nm and 500-5000~nm wide slits on the figure 4(a)). From the figure, one can conclude that the spatial limit of effective coupling for this irradiation dose is between 200~nm and 500~nm, and therefore the effective coherence length in damaged YBCO (i.e. effective $\xi_N$) is about 100-250~nm. Beyond this value the damaged segment behaves as "bulk" superconductor of reduced critical temperature (T$_{c}^{\prime}$) and T$_J$ is reduced to T$_{c}^{\prime}$.\par
For our samples, the total length of a 4 JJ array and of a 8 JJ array were 400~nm and 880~nm respectively. For the 4-JJ samples we observed sound Shapiro steps indicating that Josephson coupling was dominating effect for broad temperature range. For the 8-JJ array, steps could be seen (inset to figure 4(b)), though traits of flux-flow regime were more pronounced especially when temperature was comparatively low.\par
At this stage the exact reason for such a long range of Josephson coupling in multi-slit systems is not fully understood. Taking into account results from \cite{chen2005} where edge-to-edge separation was 75~nm we suggest that we were close to multi-junction regime. However in our case, unlike \cite{chen2005}, undamaged islands were not big enough to separate junctions from each other (probably because of 70~nm thick films we used instead of 200~nm in \cite{chen2005}). At the same time, the presence of the islands may have boosted Josephson coupling effect. In this context, we would like to mention certain similarity of our situation with "giant" Josephson coupling observed along c-axis in cuprates (\cite{bozovic2004} and references there). Theoretical approach to describe the phenomena was suggested {\cite{kresin2003,kresin2006}}. The main idea is that two distanced superconductors (Gor'kov pairing functions F$_L$ and F$_R$ do not overlap) separated with a chain of superconducting islands (with they own pairing functions F$_1$, F$_2$...F$_N$) might form a single Josephson junction if F$_L$ overlaps with F$_1$, F$_1$ with F$_2$... and F$_N$ with F$_R$. Deeper theoretical investigation remains necessary to explain all aspects of experimental results.  
\section{Conclusion}
Josephson mixing on giant Shapiro steps of ion-irradiated Josephson arrays has been demonstrated. Replacing single junction by an array would allow us to improve impedance mismatch of previously reported Josephson mixers (\cite{malnou2012,malnou2014}). To improve noise temperature of the mixer as well as for many other possible applications mutual phase locking between junctions in an array is necessary. We believe that for this particular type of junctions short-range interaction may be a promising approach to achieve synchronization in arrays. \par
Our first experiments showed that 80~nm-spaced junctions behave as single weak links. It offers an interesting possibility to increase I$_{c}$R$_{N}$ product of ion-irradiated junctions: not optimized single-slit junctions we used for the reported arrays had a normal resistance R$_{N}\approx$1$\Omega$ when the critical current was I$_{c}\approx$25$\mu$A and 8-slit junction demonstrated R$_{N}\approx$3$\Omega$ for the same I$_{c}$.\par
Properties of closely spaced arrays are therefore of high interest both from fundamental and practical points of view but remain to be investigated in details.    
 
\section*{Acknowledgments}
The authors thank Yann Legall for ion irradiations and Thomas Wolf for picture 4(a). Useful comments of V. Kresin, Yu. Ovchinnickov and V. Kornev are greatly appreciated. This work has been supported by the T-SUN ANR ASTRID program (ANR-13-ASTR-0025-01), the Emergence program Contract of Ville de Paris and by the R\'{e}gion Ile-de-France in the framework of the DIM Nano-K and Sesame programs.
\bibliographystyle{ieeetr} 
\bibliography{arrays_paper}
\end{document}